\documentclass[10pt,
nofootinbib,showpacs,reprint,
aps,prd]{revtex4-1}
\usepackage{amsfonts,amsmath,amssymb,mathrsfs}
\usepackage{mathtools,mathptmx,array,url}
\usepackage{graphics,graphicx}
\usepackage[colorlinks=true,linkcolor=blue,citecolor=blue,urlcolor=blue]{hyperref}

\def\be{\begin{equation}}
\def\ee{\end{equation}}
\def\bea{\begin{eqnarray}}
\def\eea{\end{eqnarray}}
\def\beal{\begin{aligned}}
\def\eeal{\end{aligned}}
\def\nn{\nonumber}
\def\p{\partial}
\def\cA{\mathcal{A}}

\begin{document}

\title{Null hypersurface caustics for high dimensional superentropic black holes}

\author{Di Wu$^{1,2}$}
\email{wdcwnu@163.com}

\author{Puxun Wu$^{1}$}
\email{pxwu@hunnu.edu.cn}

\affiliation{$^{1}$Department of Physics and Synergetic Innovation Center for
Quantum Effect and Applications, Hunan Normal University, Changsha, Hunan 410081,
People's Republic of China \\
$^{2}$College of Physics and Space Science, China West Normal University, Nanchong,
Sichuan 637002, People's Republic of China}

\date{\today}

\begin{abstract}
A black hole is superentropic if it violates the reverse isoperimetric inequality. Recently,
some studies have indicated that some four dimensional superentropic black holes have the
null hypersurface caustics (NHC) outside their event horizons. In this paper, we extend to
explore whether or not the NHC exists in the cases of high dimensional superentropic black
holes. We consider the singly rotating Kerr-AdS superentropic black holes in arbitrary
dimensions and the singly rotating charged superentropic black hole in five-dimensional
minimal gauged supergravity, and find that the NHC exists outside the event horizons for
these superentropic black holes. Furthermore, the spacetime dimensions and other parameters
of black hole, such as the electric charge, have important impact on the NHC inside the
horizon. Our results indicate that when the superentropic black hole has the Cauchy horizon,
the NHC will also exist inside the Cauchy horizon.
\end{abstract}


\maketitle

\section{Introduction}

Black hole is one of the most remarkable and quite fascinating objects predicted by the Einstein's
general relativity. It has been listened by the LIGO-Virgo observation \cite{PRL116-061102,
PRL116-241103} and seen through the Event Horizon Telescope \cite{APJ875-L1,APJ875-L2}, which
spurs an increasing investigation on the properties of all kinds of black hole. Among them,
the asymptotically anti de-Sitter (AdS) black hole becomes particular interest since it is very
significant in various gauge-gravity dualities. Almost ten years ago, Cveti\v{c} et al. conjectured
that the AdS black hole  satisfies the reverse isoperimetric inequality (RII) \cite{PRD84-024037}
\bea\label{RII}
\mathcal{R} = \Big[\frac{(d-1)V}{\cA_{d-2}}\Big]^{1/(d-1)}\Big(\frac{\cA_{d-2}}{A}\Big)^{1/(d-2)}
\ge 1 \, ,
\eea
where $V$ is the thermodynamic volume of black hole, $\cA_{d-2} = 2\pi^{[(d-1)/2]}/\Gamma[(d-1)/2]$
is the area of the unit $(d-2)$-sphere with $d$ being the number of spacetime dimensions and $A$ is
the area of the outer horizon. Equality is attained for the Schwarzschild-AdS black hole, which
indicates that the Schwarzschild-AdS black hole has the maximum entropy. In other words, it implies
that for a specified entropy, the Schwarzschild-AdS black hole occupies the least volume.

However, if taking the ultraspinning limit for the rotating  AdS black holes, which leads to that
black holes have finite area but noncompact event horizons, Hennigar et al.~\cite{PRL115-031101}
found that the area entropy of ultraspinning Kerr-Newman-AdS$_4$ black hole \cite{PRD89-084007,
JHEP0114127} exceeds the maximum entropy limit since $\mathcal{R} < 1$. Thus, this kind of black
holes is dubbed ``superentropic". The superentropic black hole is the first example violating
the RII. Soon afterward, a lot of new ultraspinning AdS black hole solutions \cite{JHEP0615096,
PRD95-046002,1702.03448,JHEP0118042,PRD102-044007,PRD103-044014} from some known rotating AdS
black holes have been generated successfully. The singly rotating Kerr-AdS ultaspinning black holes
in arbitrary dimensions as well as singly rotating ultraspinning black hole of five-dimensional
minimal gauged supergravity theory satisfy the relation $\mathcal{R} < 1$ \cite{JHEP0615096}.
However some new ultraspinning AdS black holes \cite{JHEP0615096,PRD102-044007,PRD103-044014}
violates the RII only in some ranges of values of the solution parameters. Thus, one naturally
has a question: what causes this new family of ultraspinning black holes to be superentropic?

Recently, it has been found that, although there is the null hypersurface caustics (NHC) only
inside the Cauchy horizon for the usual Kerr(-Newman)-(A)dS black hole~\cite{CQG36-245017},
after taking the ultraspinning limit the NHC can  exist outside the event horizon of the
Kerr(-Newman)-AdS superentropic black hole \cite{CQG38-045018}. The existence of NHC means
that the causal structure of spacetime has some pathologies. The work in Ref. \cite{CQG38-045018}
gives rise to an indication: the presence of the NHC outside the event horizon may be related
with the superentropy. Subsequently, the NHC of ultraspinning Kerr-Sen-AdS$_4$ black hole was
studied in Ref. \cite{PRD103-024053}, and it was found that for the ultraspinning Kerr-Sen-AdS$_4$
black hole, whether it is superentropic or not, the NHC always appears both out and inside of
the horizon. Obviously, the existence of NHC outside the horizon is only investigated for several
four dimensional superentropic black holes, and whether the high dimensional superentropic black
holes have the NHC outside their horizons or not needs to be explored, which motivates us to
undertake the present work.

In this paper, we investigate the NHC for two high dimensional superentropic black holes, including
the singly rotating Kerr-AdS superentropic black holes in arbitrary dimensions \cite{PRL115-031101}
and the singly rotating charged superentropic black hole in five-dimensional minimal gauged
supergravity theory \cite{JHEP0615096}. We find that for these superentropic black holes, there
is NHC outside the event horizon. But the presence of NHC inside the horizon seems to require
the existence of Cauchy horizon. The remaining part of this paper is organized as follows. Sec.
\ref{II} is a brief review of the NHC of the Kerr-AdS$_4$ superentropic black hole. In Sec. \ref{III}
and Sec. \ref{IV}, we investigate the NHC of the singly rotating Kerr-AdS superentropic black holes
in arbitrary dimensions and the singly rotating charged superentropic black hole in five-dimensional
minimal gauged supergravity theory, respectively. Finally, our conclusions are given in Sec. \ref{V}.

\section{The NHC of Kerr-AdS$_4$ superentropic black hole}\label{II}

For the four-dimensional Kerr-AdS superentropic black hole, its metric has the form
\cite{PRL115-031101,JHEP0615096,JHEP0114127,PRD89-084007}:
\bea
ds^2 &=& -\, \frac{\Delta_r}{\Sigma}\big(dt -l\sin^2\theta d\phi\big)^2
 +\Sigma\left(\frac{dr^2}{\Delta_r} +\frac{d\theta^2}{\sin^2\theta}\right) \nn \\
&& +\frac{\sin^4\theta}{\Sigma}\left[l\, dt -(r^2 +l^2)d\phi\right]^2 \, , \label{SEKNAdS}
\eea
where $l$ is the cosmological scale, and
$$ \Delta_r = \left(r^2 +l^2 \right)^2/l^2 -2mr \, , \quad~~ \Sigma = r^2 +l^2\cos^2\theta \, . $$
Here, $m$ is the mass parameter. The azimuthal coordinate $\phi$ is noncompact and must be
compactified by requiring $\phi \sim \phi +\mu$ with $\mu$ being a dimensionless parameter
related to a new chemical potential $K$.

The horizon of  black hole satisfies the equation
\be\label{hcSEKNAdS4}
\Delta_r = 0 \, .
\ee
It is worth to note that the existence of the inner and outer horizons requires the mass to be
larger than a critical value $m_e$
\be
m_e \equiv 2r_e\left(\frac{r_e^2}{l^2} +1\right)  = \frac{8l}{3\sqrt{3}}\, ,
\ee
where
\be
r_e = \frac{l}{\sqrt{3}} \, . \nn
\ee
When $m = m_e$, the superentropic Kerr-AdS$_4$ black hole is extremal. While for $m < m_e$, it
has a naked singularity.

Since outside the black hole horizon $g_{\phi\phi}$ satisfies
\bea\label{gpp}
g_{\phi\phi} &=& \frac{2mrl^4\sin^4\theta}{\Sigma}\ge 0 \, ,
\eea
and thus is strictly positive, the spacetime is free of closed timelike curves, which is
consistent with the result given in Ref. \cite{JHEP0615096}.

Now, we turn to investigate the NHC for superentropic Kerr-AdS$_4$ black hole. Following
Refs.~\cite{CQG15-2289,CQG36-245017,CQG38-045018,PRD103-024053}, we introduce the outgoing
and ingoing Eddington-Finkelstein coordinates defined in terms of the ``generalized tortoise
coordinate" $r_*(r,\theta)$:
\be\label{uv1}
u = t - r_*(r,\theta) \, , \qquad v = t +r_*(r,\theta) \, .
\ee
The null hypersurfaces are described by
\be\label{uv}
u = const \, , \qquad v = const  \, ,
\ee
which are dubbed the outgoing and ingoing null congruences of the hypersurfaces, respectively.
It is easy to obtain that  the null hypersurfaces defined by $u = const$ or $v = const$ satisfy
the equation
\be\label{PDE}
g^{\mu\nu}\p_\mu(t +r_*)\p_\nu(t +r_*) = g^{tt} +g^{rr}(\p_r r_*)^2
 +g^{\theta\theta}(\p_\theta r_*)^2 = 0 \, .
\ee
Solving this partial differential equation can give the null hypersurfaces for the superentropic
Kerr-AdS$_4$ metric (\ref{SEKNAdS}).

Using the contravariant components $g^{tt}$,  $g^{rr}$, and $g^{\theta\theta}$ of the metric
(\ref{SEKNAdS}), the partial differential equation (\ref{PDE}) can be re-expressed as
\be\label{PDE2}
l^2 -\frac{\left(r^2 +l^2 \right)^2}{\Delta_r} +\Delta_r(\p_r r_*)^2
 +\sin^2\theta(\p_\theta r_*)^2 = 0 \, .
\ee
Introducing the so-called ``constant of separation" $l^2\lambda$ (hereinafter, $\lambda$ is
referred to as the separation constant), from Eq.~(\ref{PDE2}) one can have
\be\label{rr2}
(\p_r r_*)^2 = \frac{Q^2(r)}{\Delta_r^2} \, , \quad (\p_\theta r_*)^2
 = \frac{P^2(\theta)}{\sin^4\theta} \, .
\ee
where
\be\beal\label{Q2P2}
&Q^2(r) = \left(r^2 +l^2 \right)^2 -l^2\lambda\Delta_r \, ,  \\
&P^2(\theta) = (\lambda -1)l^2\sin^2\theta \,
\eeal\ee
with $\lambda > 1$. Here a complex $\theta-$dependent form of $P(\theta)$ is chosen in order to
give the elegant expression for the following equation (\ref{nu}). Using  the relations given in
(\ref{rr2}), one can express the total differential $dr_* = \p_r r_*dr +\p_\theta r_*d\theta$ to be
\be\label{dr}
dr_* = \frac{Q(r)}{\Delta_r}dr +\frac{P(\theta)}{\sin^2\theta}d\theta \, .
\ee

If treating $\lambda$ as a variable,  the exact differential Eq.~(\ref{dr}) can be generalized to be
\be\label{dr1}
dr_* = \frac{Q(r,\lambda)}{\Delta_r}dr +\frac{P(\theta,\lambda)}{\sin^2\theta}d\theta
 +c_1F(r,\theta,\lambda)d\lambda \, ,
\ee
where $c_1$ is an arbitrary constant and $F(r,\theta,\lambda)$ is an arbitrary function. Since
Eqs.~(\ref{dr}) and (\ref{dr1}) are functionally equivalent, the condition
\be
F(r,\theta,\lambda) = 0
\ee
needs to be satisfied,  which indicates $dF(r,\theta,\lambda) = 0$ and thus yields
\be\label{dF}
[\p_\lambda F(r,\theta,\lambda)]d\lambda +[\p_r F(r,\theta,\lambda)]dr
 +[\p_\theta F(r,\theta,\lambda)]d\theta
= 0 \, .
\ee
From the Poincar\'e lemma in the external differentiation theory, namely $d(dr_*) = 0$, one can
obtain the following integrable conditions
\be\beal
&\frac{\p_\lambda Q(r,\lambda)}{\Delta_r} = c_1\p_rF(r,\theta,\lambda)\, , \\
&\frac{\p_\lambda P(\theta,\lambda)}{\sin^2\theta} = c_1\p_\theta F(r,\theta,\lambda) \, .
\eeal\ee
From the definitions of $Q(r, \lambda)$ and $P(\theta, \lambda)$, one can obtain straightforwardly
\be
\p_\lambda Q(r,\lambda) = -\frac{l^2\Delta_r}{2Q(r,\lambda)} \, , \quad
 \p_\lambda P(\theta,\lambda) = -\frac{l^2\sin^2\theta}{2P(\theta,\lambda)} \, ,
\ee
and then rewrite Eq. (\ref{dF}) as
\be\label{nu}
\nu d\lambda = -\frac{dr}{Q(r,\lambda)} +\frac{d\theta}{P(\theta,\lambda)} \, ,
\ee
after choosing the constant $c_1$ to be $l^2/2$ and defining $\nu = -\p_\lambda F(r,\theta,\lambda)$.
Making use of Eq. (\ref{dr}) and Eq. (\ref{nu}), the metric of the superentropic Kerr-AdS$_4$ metric
given in (\ref{SEKNAdS}) can be re-expressed in terms of the coordinates ($t,r_*,\theta,\lambda$)
\bea
ds^2 &=& \frac{\Delta_r\sin^2\theta}{R^2}(dr_*^2 -dt^2) +R^2\sin^2\theta(d\phi -\Omega dt)^2 \nn \\
&&+\frac{\nu^2P^2(\theta,\lambda)Q^2(r,\lambda)}{R^2}d\lambda^2 \, ,
\eea
where
\bea
R^2 &=& \frac{g_{\phi\phi}}{\sin^2\theta} = \frac{2mrl^2\sin^2\theta}{\Sigma}  \, , \\
\Omega &=& -\frac{g_{t\phi}}{g_{\phi\phi}} = \frac{(r^2 +l^2)\sin^2\theta -\Delta_r}{\Sigma R^2} \, .
\eea
Obviously, $\Omega$ is the Zero-Angular-Momentum-Observer (ZAMO) angular velocity of the
superentropic Kerr-AdS$_4$ black hole \cite{PRL115-031101,JHEP0615096,PRD101-024057}. Then,
using  Eqs.~(\ref{dr}) and (\ref{nu}), one has
\bea
&&dr = \frac{Q(r,\lambda)\Delta_r}{\Sigma R^2}\left[\sin^2\theta dr_*
 -P^2(\theta,\lambda)\nu d\lambda \right] \, , \label{dr2} \\
&&d\theta = \frac{P(\theta,\lambda)\sin^2\theta}{\Sigma R^2}\left[\Delta_r dr_*
 +Q^2(r,\lambda)\nu d\lambda \right] \, . \label{dtheta}
\eea

Since the outgoing and ingoing null congruences are defined to be $du = dv = 0$, which means
$dr_*^2 = dt^2$, the metric of the superentropic Kerr-AdS$_4$ black hole on the null hypersurface
is reduced to be
\be\label{dh2}
dh^2 = R^2\sin^2\theta(d\phi -\Omega dt)^2 +\frac{\nu^2P^2(\theta,\lambda)
 Q^2(r,\lambda)}{R^2}d\lambda^2\, .
\ee
One can see that the null congruences are rotating with the ZAMO angular velocity $\Omega$.
Since that the volume element of a null hypersurface is the square root of the determinant of
the induced metric, the points where the induced metric determinant goes to zero correspond
to the NHC. Thus, the condition of the NHC is
\be\label{NHC}
\nu P(\theta,\lambda)Q(r,\lambda)\sin\theta =\nu l\sqrt{\lambda-1}Q(r,\lambda) \sin^2\theta = 0 \, ,
\ee
which is the determinant of  the induced metric of the superentropic Kerr-AdS$_4$ black hole.

Let us consider the outgoing null hypersurface case  (same for ingoing null hypersurface case),
namely, $\lambda = const$ and a increasing $r$, as an example to study NHC.  We need to analyze
each factor in Eq. (\ref{NHC}). First of all, $l$ is the AdS radius and assumed to be always
positive. According to Eq.~(\ref{dtheta}), $\theta$ can be expressed as $\theta(r,\lambda)$.
Combining  Eqs.~(\ref{dr2}) and (\ref{dtheta}) yields
\be
\left(\frac{\p \theta}{\p r}\right)_{\lambda} = \frac{P(\theta)}{Q(r)} \ge 0 \, ,
\ee
which means that $\theta$ increases, or remains to be a constant, as the radius of the superentropic Kerr-AdS$_4$ black hole increases. According to Eq.~(\ref{nu}), one can easily obtain
\bea
&&\frac{\p\nu}{\p r} = -\frac{\p}{\p\lambda}\frac{\p F(r,\theta,\lambda)}{\p r}
 = \frac{l^2\Delta_r}{2Q^3(r,\lambda)}\, , \\
&&\frac{\p\nu}{\p\theta} = -\frac{\p}{\p\lambda}\frac{\p F(r,\theta,\lambda)}{\p\theta}
 = \frac{l^2\sin^2\theta}{2P^3(\theta,\lambda)} \, .
\eea
Hence, we have
\bea\label{nur}
\left(\frac{\p\nu}{\p r} \right)_\lambda = \frac{\p\nu}{\p r} +\frac{\p\nu}{\p\theta}
 \left(\frac{\p\theta}{\p r} \right)_\lambda = \frac{mrl^2}{(\lambda -1)Q^3(r)} \, .
\eea
Since $\lambda > 1$, Eq. (\ref{nur}) is always positive, which means that once the radius of the superentropic Kerr-AdS$_4$ black hole increases, so does $\nu$. When $r\rightarrow \infty$,
$\nu$ is not divergent since $({\p\nu}/{\p r} )_\lambda\rightarrow r^{-5}$. Clearly, if  $Q(r) = 0$,
Eq.~(\ref{NHC}) is always satisfied. Thus, $Q(r)=0$ is a sufficient condition that leads to the
existence of NHC.  $Q(r) = 0$ means that
\be\label{ccSEKNAdS4}
\left(r^2 +l^2 \right)^2 -l^2\lambda\Delta_r = 0,
\ee
which is a quartic equation of $r$ and is hard to give the analytical results.

\begin{figure}[ht!]
\centering
\includegraphics[width=0.45\textwidth]{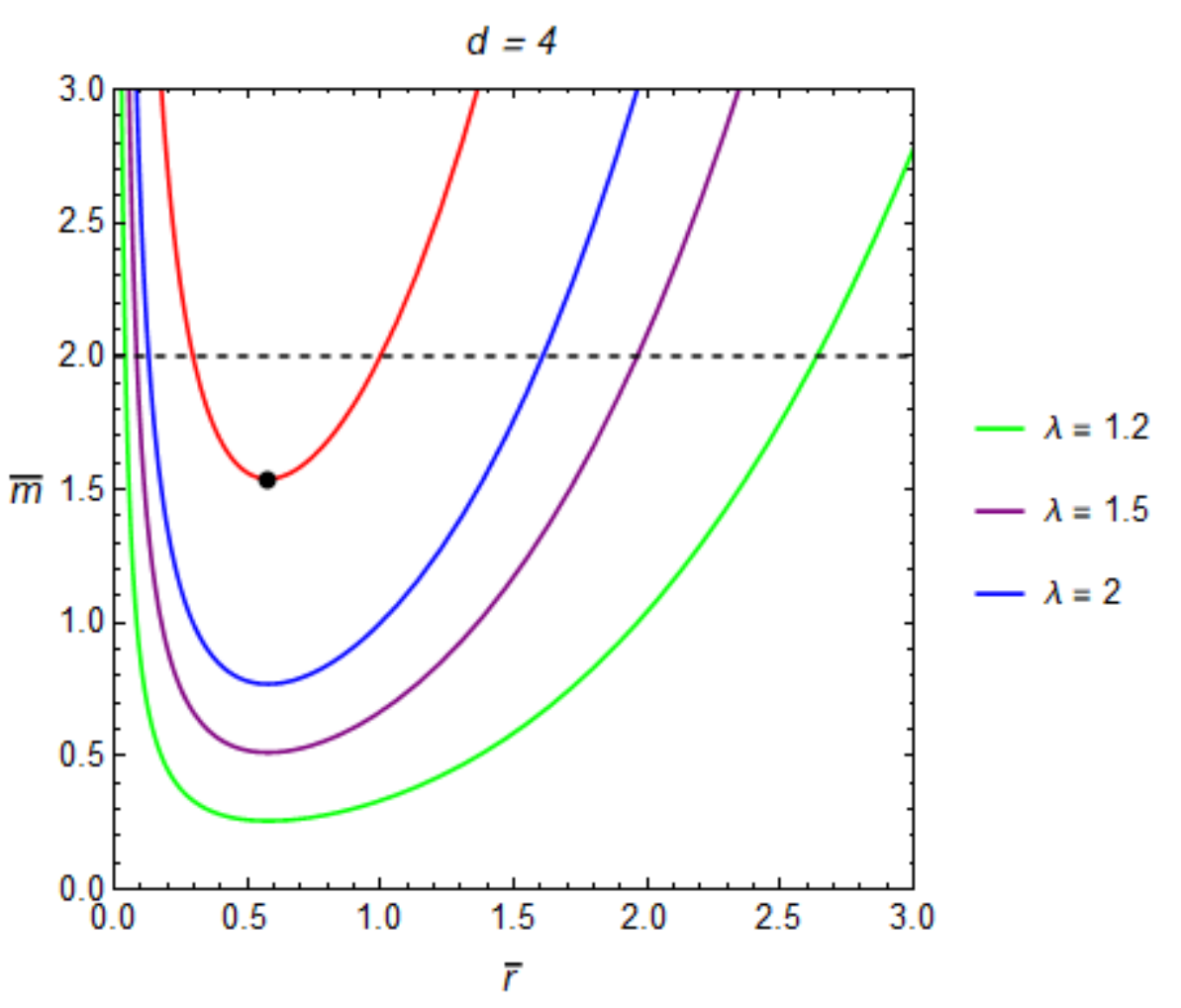}
\caption{Contours of the caustics on the ($\bar{r}$,$\bar{m}$)-plane with different values
of $\lambda$ for the superentropic Kerr-AdS$_4$ black hole, where $\bar{m} = m/l$ and $\bar{r}
= r/l$. The red line is the horizon equation and the black dot represents the extremal black
hole case.
\label{SEKAdS4NHC}}
\end{figure}

In FIG.~\ref{SEKAdS4NHC}, we show the numerical results. In this figure the horizon equation
($\Delta_r=0$) is also plotted with the solid red line, and the black dot represents the
extremal black hole case. Two intersection points between the dashed black line and the solid
red line represent the Cauchy horizon and the event horizon of the black hole, respectively.
So, it is easy to find that the NHC appears both inside the Cauchy horizon and outside the
event horizon of the superentropic Kerr-AdS$_4$ black hole, which is consistent with the
conclusion given in Ref. \cite{CQG38-045018}.

\section{Singly rotating Kerr-AdS superentropic black holes in arbitrary dimensions}\label{III}

In this section, we will extend the above discussion to the case of superentropic black hole
with higher dimensions by considering the arbitrary dimensional singly rotating Kerr-AdS
superentropic black holes. The singly rotating Kerr-AdS superentropic black holes in arbitrary
dimensions are strictly superentropic \cite{JHEP0615096} and the metric has the form
\cite{PRL115-031101,PRD101-024057}
\be
\beal\label{SEKAdS}
d\bar{s}^2 &= -\, \frac{\bar{\Delta}_r}{\bar\Sigma}\left(dt -l\sin^2\theta d\phi\right)^2
 +\frac{\bar{\Sigma}}{\bar{\Delta}_r}dr^2 \\
&\quad +\frac{\bar\Sigma}{\sin^2\theta}d\theta^2
 +\frac{\sin^4\theta}{\bar\Sigma}\left[ldt -(r^2 +l^2)d\phi\right]^2 \\
&\quad +r^2\cos^2\theta\, d\Omega_{d-4}^2 \, ,
\eeal
\ee
where
\be
\bar{\Delta}_r = \left(r^2 +l^2 \right)^2/l^2 -2mr^{5-d} \, , \quad
\bar\Sigma = r^2 +l^2\cos\theta \, , \nn
\ee
and the horizon equation is
\be\label{hcSEKAdS}
\bar{\Delta}_r = 0 \, .
\ee

As did in the last section, we can derive that the null hypersurfaces equation is
\be\label{SEKAdSNHE}
l^2 -\frac{\left(r^2 +l^2 \right)^2}{\bar{\Delta}_r} +\bar{\Delta}_r(\p_r r_*)^2
 +\sin^2\theta(\p_\theta r_*)^2 = 0 \, .
\ee
Introducing a constant $\lambda$ with $\lambda>1$ for the separation of variables, i.e.
$l^2\lambda$, the above equation (\ref{SEKAdSNHE}) yields
\be\beal
&(\p_r r_*)^2 = \frac{\left(r^2 +l^2 \right)^2 -l^2\lambda\bar{\Delta}_r}{\bar{\Delta}_r^2} \, , \\
&(\p_\theta r_*)^2 = \frac{(\lambda -1)l^2\sin^2\theta}{\sin^4\theta} \, .
\eeal\ee
After a similar procedure in the previous section, one can find that  the caustics condition is
\be\label{ccSEKAdS}
\left(r^2 +l^2 \right)^2 -l^2\lambda\bar{\Delta}_r = 0 \, .
\ee

\begin{figure}[ht!]
\centering
\includegraphics[width=0.45\textwidth]{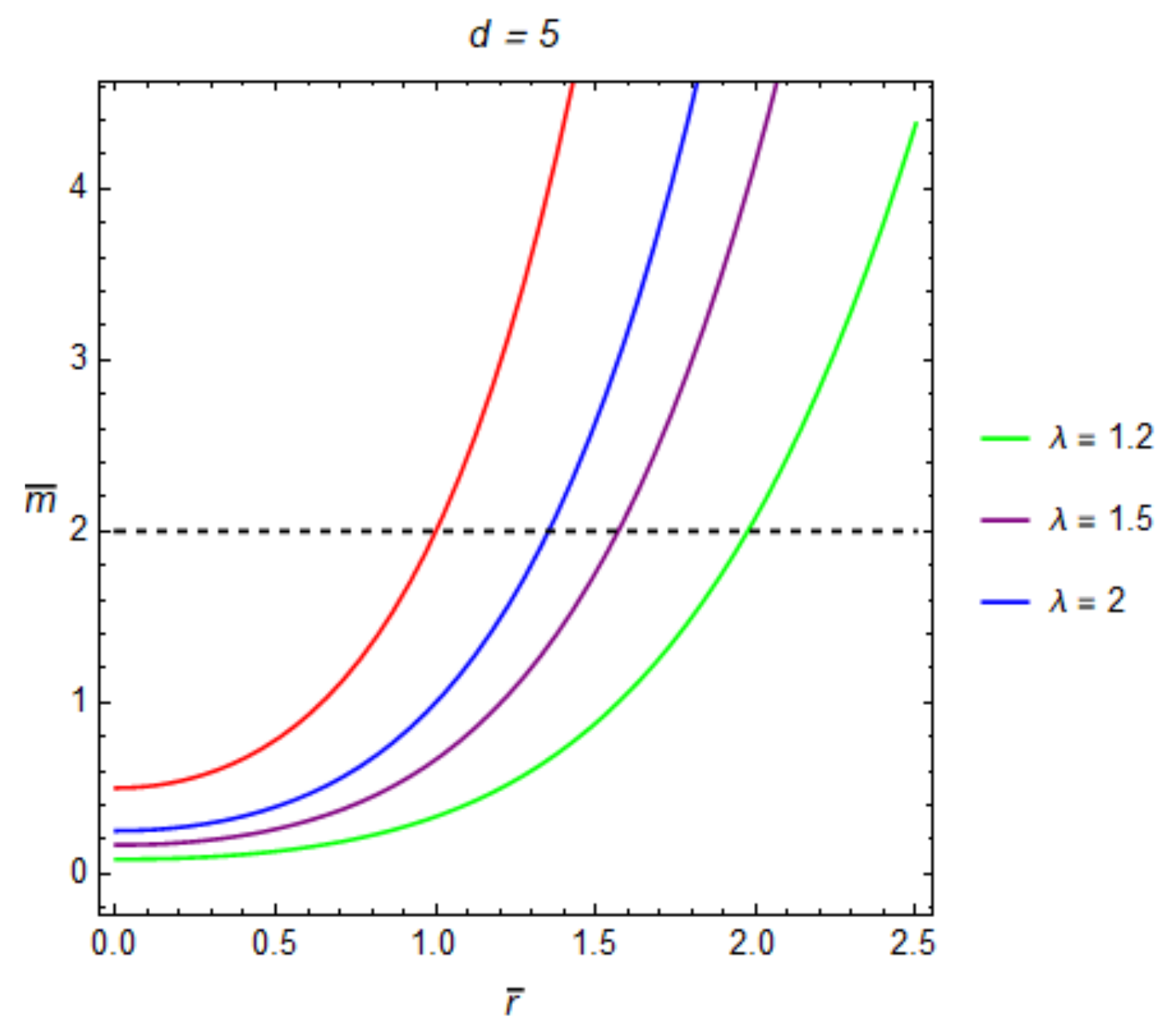}
\caption{Contours of the caustics on the ($\bar{r}$,$\bar{m}$)-plane for the singly rotating
Kerr-AdS$_5$ superentropic black hole, where $\bar{m} = m/l^2$ and $\bar{r} = r/l$. The red
line is the horizon eqaution.
\label{SEKAdS5NHC}}
\end{figure}

To study the effect of dimensions on the caustics condition (\ref{ccSEKAdS}) and the horizon
equation (\ref{hcSEKAdS}), we consider the five and six dimensions, respectively, as examples.
For singly rotating Kerr-AdS$_5$ superentropic black hole, one can obtain that the caustics
condition and horizon equation become
\bea
&&\left(r^2 +l^2\right)^2 -l^2\lambda\left[\left(r^2 +l^2 \right)^2/l^2
 -2m \right] = 0 \, , \label{dccSEKAdS5} \\
&&\left(r^2 +l^2 \right)^2/l^2 -2m = 0 \, . \label{dhcSEKAdS5}
\eea
These two identifies are shown in FIG.~\ref{SEKAdS5NHC} with the horizon equation being represented
by the red solid line. One can see that the black dashed line and the red line only has one
intersection point when $\bar{r} > 0$, which means that the singly rotating Kerr-AdS$_5$
superentropic black hole only has the event horizon and the Cauchy horizon vanishes. This
can also be known from  Eq. (\ref{dhcSEKAdS5}) since it has only one positive real root
$r = \sqrt{-l^2 +l\sqrt{2m}}$. This property is the same as the singly rotating Kerr-AdS$_5$
black hole \cite{PRD59-064005}, but is different from the case of the superentropic Kerr-AdS$_4$
black hole discussed in the above section. Furthermore, comparing FIG.~\ref{SEKAdS5NHC} with
FIG.~\ref{SEKAdS4NHC}, we find that the NHC inside the horizon vanishes, and the NHC only exists
outside the event horizon when the spacetime dimension  is increased from four to five.

\begin{figure}[ht!]
\centering
\includegraphics[width=0.45\textwidth]{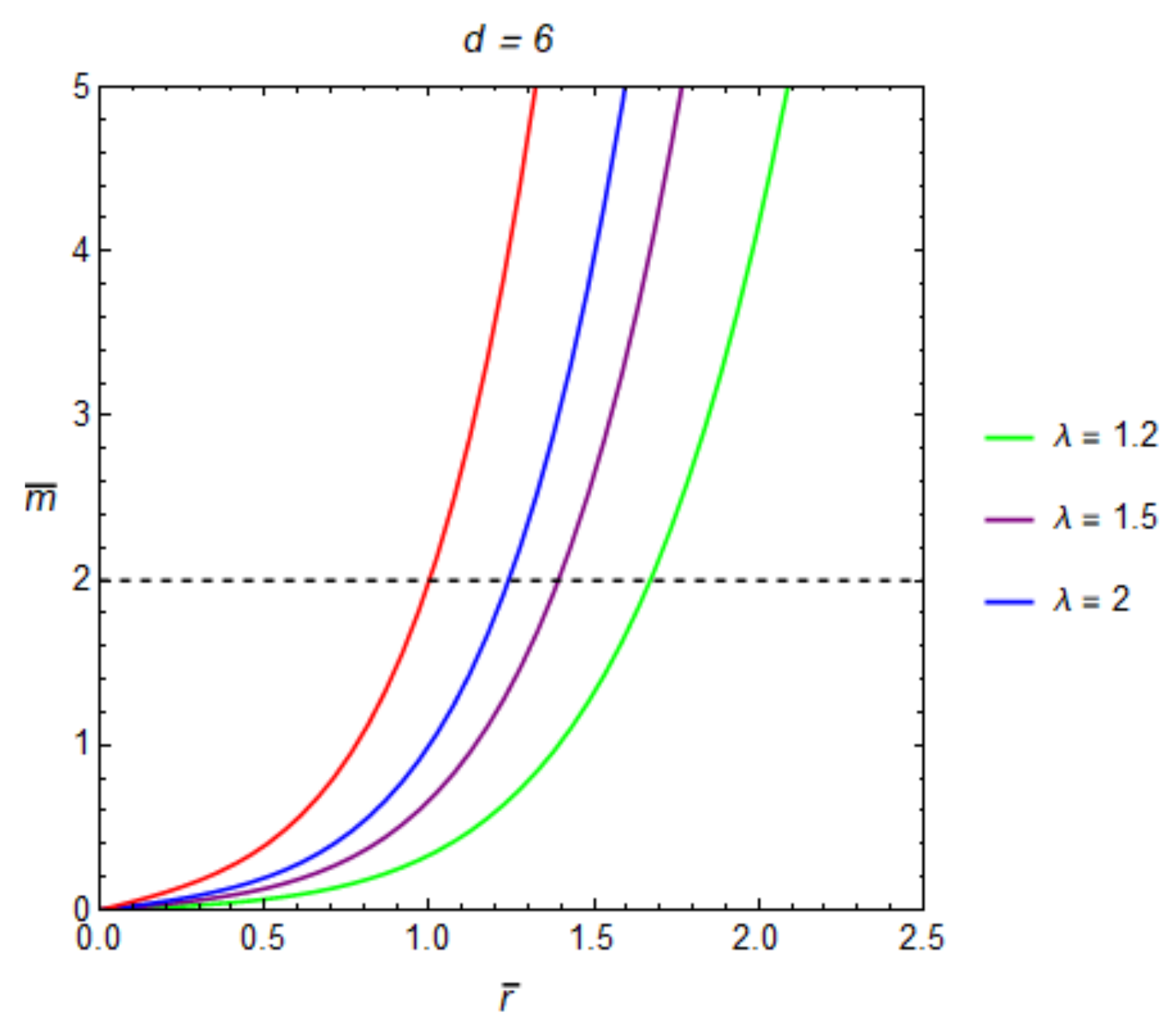}
\caption{Contours of the caustics  on the ($\bar{r}$,$\bar{m}$)-plane for the singly rotating
Kerr-AdS$_6$ superentropic black hole, where $\bar{m} = m/l^3$ and $\bar{r} = r/l$. The red
line is the horizon equation.
\label{SEKAdS6NHC}}
\end{figure}

For the singly rotating Kerr-AdS$_6$ superentropic black hole, its caustics and horizon equations,
respectively, satisfy
\bea
&&\left(r^2 +l^2 \right)^2 -l^2\lambda\left[\left(r^2 +l^2 \right)^2/l^2
 -2m/r \right] = 0 \, , \label{dccSEKAdS6} \\
&&\left(r^2 +l^2 \right)^2/l^2 -2m/r  = 0 \, , \label{dhcSEKAdS6}
\eea
which are plotted in FIG.~\ref{SEKAdS6NHC}. In this figure, the horizon equation is shown by the
red solid line.  FIG.~\ref{SEKAdS6NHC} is very similar to FIG.~\ref{SEKAdS5NHC}, and the same
conclusion that the NHC only exists outside the event horizon of superentropic black hole can be
obtained. Thus, increasing the spacetime dimension from five to six has negligible effect on the
NHC. Comparing FIG.~\ref{SEKAdS4NHC}, FIG.~\ref{SEKAdS5NHC} and FIG.~\ref{SEKAdS6NHC}, one can
infer that the existence of the NHC outside the event horizon of superentropic black hole, but
once the spacetime dimension is larger than four, then the Cauchy horizon of superentropic black
hole disappears and the NHC existing inside the horizon  will  disappear too. Therefore, the
dimension of spacetime only has an important effect on the property of NHC inside the horizon.

\section{Singly rotating charged superentropic black hole in five-dimensional minimal gauged
supergravity}\label{IV}

To investigate whether the conclusion obtained in Sec. \ref{III} is general or not, in this
section we turn to analyze the NHC of the singly rotating charged superentropic black hole
in five-dimensional minimal gauged supergravity theory. According to the solution of the
double rotating charged superentropic black hole in five-dimensional minimal gauged supergravity
theory~\cite{JHEP0615096}, we obtain the singly rotating charged superentropic black hole
solution by taking one of the angular velocity, which is not boosted to the speed of light,
to be zero. Then the metric and the Abelian gauged potential are
\bea
d\hat{s}^2 &=& -\frac{\hat{\Delta}_r}{\hat{\Sigma}}\left(dt -l\sin^2\theta d\phi\right)^2
 +\frac{\hat{\Sigma}}{\hat{\Delta}_r}dr^2 \nn \\
&&+\frac{\hat{\Sigma}}{\sin^2\theta}d\theta^2
+\frac{\sin^4\theta}{\hat{\Sigma}}\left[ldt -(r^2 +l^2)d\phi\right]^2 \nn \\
&&+r^2\cos^2\theta\left[d\psi -\frac{ql(dt
  -l\sin^2\theta d\phi)}{r^2\hat{\Sigma}}\right]^2 \, , \label{SECCLP} \\
\hat{A} &=& \frac{\sqrt{3}q}{2\hat{\Sigma}}(dt -l\sin^2\theta d\phi) \, ,
\eea
where
\be
\hat{\Delta}_r = \left(r^2 +l^2 \right)^2/l^2 -2m +q^2/r^2 \, , \quad
\hat{\Sigma} = r^2 +l^2\cos^2\theta \, , \nn
\ee
with $q$ being the electric charge parameter. The horizon equation is
\be\label{hcSECCLP}
\hat{\Delta}_r = 0 \, .
\ee
Obviously, it can reduce consistently that of the singly rotating Kerr-AdS$_5$ superntropic
black hole given in (\ref{SEKAdS}) when  $q = 0$.

To discuss the NHC,  we first obtain the null hypersurfaces equation
\be\label{SECCLPNHE}
l^2 -\frac{\left(r^2 +l^2 \right)^2}{\hat{\Delta}_r} +\hat{\Delta}_r(\p_r r_*)^2
 +\sin^2\theta(\p_\theta r_*)^2 = 0 \, .
\ee
After introducing a constant for the separation of variables as $l^2\lambda$ with $\lambda>1$,
the above equation (\ref{SECCLPNHE}) yields
\be\beal
&(\p_r r_*)^2 = \frac{\left(r^2 +l^2 \right)^2 -l^2\lambda\hat{\Delta}_r}{\hat{\Delta}_r^2} \, , \\
&(\p_\theta r_*)^2 = \frac{(\lambda -1)l^2\sin^2\theta}{\sin^4\theta} \, .
\eeal\ee
After a similar procedure as what was done in Sec. II, one can obtain that the caustics condition is
\be\label{ccSECCLP}
\left(r^2 +l^2 \right)^2 -l^2\lambda\hat{\Delta}_r = 0 \, .
\ee

\begin{figure}[ht!]
\centering
\includegraphics[width=0.45\textwidth]{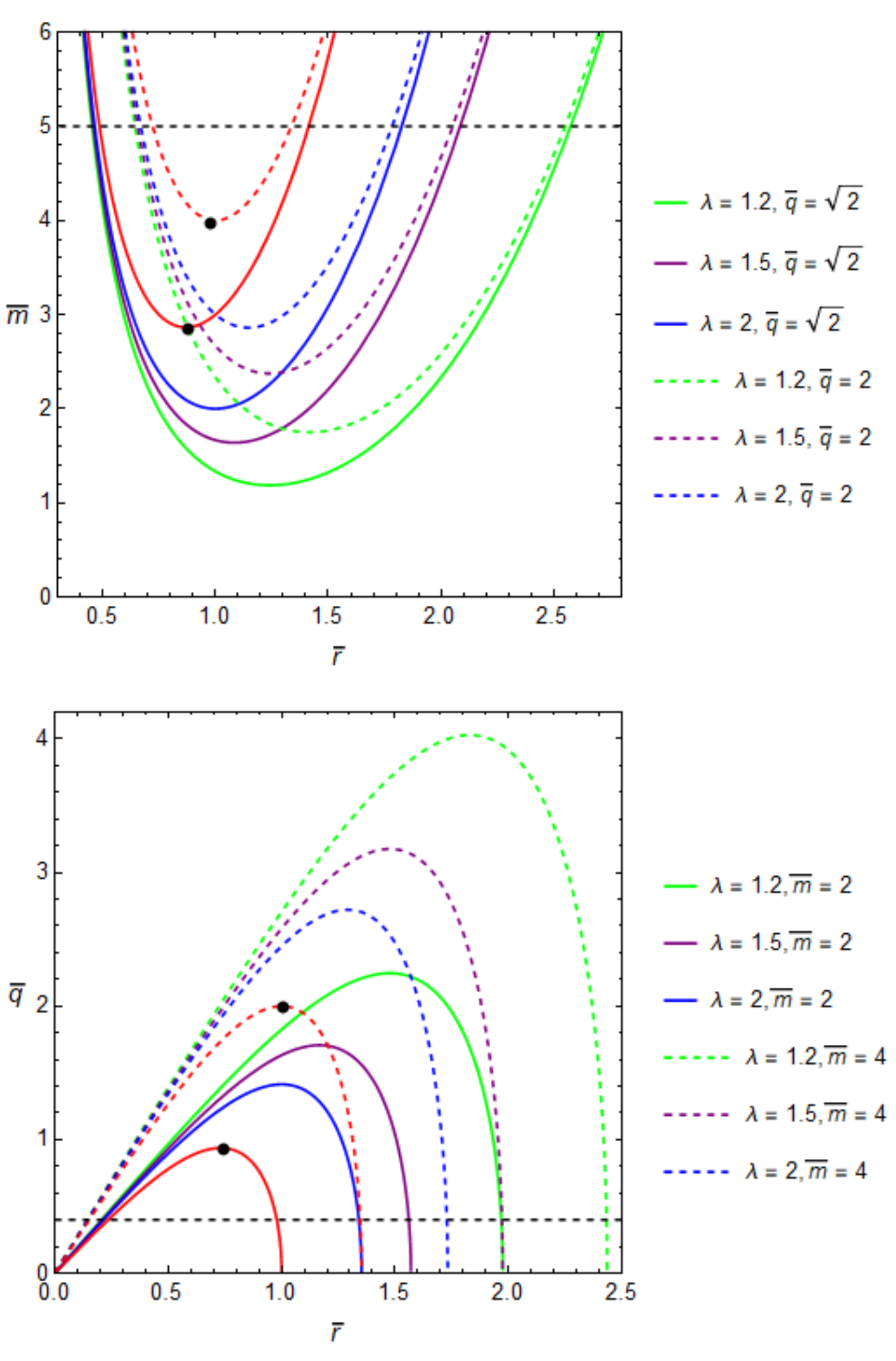}
\caption{ Contours of the caustics  on the ($\bar{r}$,$\bar{m}$)-plane with a fixed $\bar{q}$,
and on the ($\bar{r}$,$\bar{q}$)-plane with a fixed $\bar{m}$ for the singly rotating charged
superentropic black hole of five-dimensional minimal gauged supergravity, where $\bar{m} = m/l^2$,
$\bar{q} = q/l^2$ and $\bar{r} = r/l$. The red line is the horizon equation and the black dot
represents the extremal black hole case.
\label{SECCLPNHC}}
\end{figure}

FIG.~\ref{SECCLPNHC} shows the numerical results of the caustic condition (\ref{ccSECCLP}) and
the horizon equation (\ref{hcSECCLP}) (red dashed/solid  lines). In this figure, the extremal
black hole case is represented by black spots. One can see that the black dashed line and the
red dashed/solid line have two intersection points, which represent, respectively, the position
of the Cauchy and event horizons of the singly rotating charged superentropic black hole of
five-dimensional minimal gauged supergravity theory. One can see that the NHC exists both
inside its Cauchy horizon and outside its event horizon. This result is different from that
of five-dimensional black hole obtained in the Sec. \ref{III}. Thus, the electric charge of
superentropic black hole has also an important impact on the property of the NHC.

\section{Conclusions}\label{V}

The superentropic black hole, which violates the RII, has spurred an increasing deal of interest.
Recently, the authors in Refs. \cite{CQG38-045018,PRD103-024053} found that there is the NHC outside
the event horizons for some four-dimensional superentropic black holes. In this paper, we have
generalized these researches presented in \cite{CQG38-045018,PRD103-024053} to the case of high
dimensional superentropic black holes. We first consider the singly rotating Kerr-AdS superentropic
black holes in arbitrary dimensions, and focus on the cases of four, five and six dimensions as
examples. When the spacetime dimension is larger than four, the superentropic black hole has only
the event horizon while its Cauchy horizon vanishes. Only in the case of four-dimensional black hole, the NHC also
exists inside the Cauchy horizon. This result seems to indicate that the spacetime dimension has
an effect on the NHC inside the horizon. After studying the singly rotating charged superentropic
black hole in five-dimensional minimal gauged supergravity, we find that the NHC exists outside
the event horizon, and the electric charge parameter leads to the presence of NHC inside the Cauchy
horizon. Thus, the spacetime dimensions and other parameters of black hole, such as the electric
charge, have important impacts on whether the NHC exists inside the black hole horizon or not.
Our results indicate that when the superentropic black hole has the Cauchy horizon, the NHC exists
also inside this Cauchy horizon. From the results of this paper and previous works \cite{CQG38-045018,PRD103-024053}, one can find that there is the NHC outside the event horizon of the superentropic black hole, but the existence of the NHC outside the event horizon of black hole does not  mean necessarily that this black hole is superentropic. Thus, there should be some relations between  the presence of the NHC outside the event horizon and  the superentropy of black hole. However,  the origin of the superentropy of black hole still needs to be further explored.

\acknowledgments

We appreciate very much the insightful comments and helpful suggestions by anonymous referee,  and thank Prof. Hongwei Yu and Prof. Shuang-Qing Wu for helpful discussions.  This work is supported by the National Natural Science Foundation of China under Grants No. 11675130,
No. 11775077, and No. 11690034, by the National Key Research and Development Program of China
Grant No. 2020YFC2201502, and by the Science and Technology Innovation Plan of Hunan province
under Grant No. 2017XK2019.


\begin{thebibliography}{99}
\def\APJ{Astrophys. J.\,}
\def\CMP{Commun. Math. Phys.\,}
\def\CQG{Classical Quantum Gravity\,}
\def\JHEP{J. High Energy Phys.\,}
\def\PRD{Phys. Rev. D\,}
\def\PRL{Phys. Rev. Lett.\,}
\def\NPB{Nucl. Phys. B\,}
\def\PLB{Phys. Lett. B\,}
\def\MPLA{Mod. Phys. Lett. A\,}

\bibitem{PRL116-061102}
B.P. Abbott \emph{et al.} (LIGO Scientific and Virgo Collaborations),
Observation of Gravitational Waves from a Binary Black Hole Merger,
\href{http://dx.doi.org/10.1103/PhysRevLett.116.061102}
{\PRL \textbf{116}, 061102 (2016)}.

\bibitem{PRL116-241103}
B.P. Abbott \emph{et al.} (LIGO Scientific and Virgo Collaborations),
GW151226: Observation of Gravitational Waves from a 22-Solar-Mass Binary Black Hole Coalescence,
\href{http://dx.doi.org/10.1103/PhysRevLett.116.241103}
{\PRL \textbf{116}, 241103 (2016)}.

\bibitem{APJ875-L1}
The Event Horizon Telescope Collaboration,
First M87 Event Horizon Telescope Results. I. The Shadow of the Supermassive Black Hole,
\href{http://dx.doi.org/10.3847/2041-8213/ab0ec7}
{\APJ \textbf{875}, L1 (2019)}.

\bibitem{APJ875-L2}
The Event Horizon Telescope Collaboration,
First M87 Event Horizon Telescope Results. II. Array and Instrumentation,
\href{http://dx.doi.org/10.3847/2041-8213/ab0c96}
{\APJ \textbf{875}, L2 (2019)}.

\bibitem{PRD84-024037}
M. Cveti\v{c}, G.W. Gibbons, D. Kubiz\v{n}\'ak, and C.N. Pope,
Black hole enthalpy and an entropy inequality for the thermodynamic volume,
\href{http://dx.doi.org/10.1103/PhysRevD.84.024037}
{\PRD \textbf{84}, 024037 (2011)}.

\bibitem{PRL115-031101}
R.A. Hennigar, R.B. Mann, and D. Kubiz\v{n}\'ak,
Entropy Inequality Violations from Ultraspinning Black Holes,
\href{http://dx.doi.org/10.1103/PhysRevLett.115.031101}
{\PRL \textbf{115}, 031101 (2015)}.

\bibitem{PRD89-084007}
D. Klemm,
Four-dimensional black holes with unusual horizons,
\href{http://dx.doi.org/10.1103/PhysRevD.89.084007}
{\PRD \textbf{89}, 084007 (2014)}.

\bibitem{JHEP0114127}
A. Gnecchi, K. Hristov, D. Klemm, C. Toldo, and O. Vaughan,
Rotating black holes in 4d gauged supergravity,
\href{http://dx.doi.org/10.1007/JHEP01(2014)127}
{\JHEP \textbf{1401} (2014) 127}.

\bibitem{JHEP0615096}
R.A. Hennigar, D. Kubiz\v{n}\'ak, R.B. Mann, and N. Musoke,
Ultraspinning limits and super-entropic black holes,
\href{http://dx.doi.org/10.1007/JHEP06(2015)096}
{\JHEP \textbf{1506} (2015) 096}.

\bibitem{PRD95-046002}
S.M. Noorbakhsh and M. Ghominejad,
Ultra-spinning gauged supergravity black holes and their Kerr/CFT correspondence,
\href{http://dx.doi.org/10.1103/PhysRevD.95.064002}
{\PRD \textbf{95}, 046002 (2017)}.

\bibitem{1702.03448}
S.M. Noorbakhsh and M. Ghominejad,
Higher dimensional charged AdS black holes at ultra-spinning limit and their 2d CFT duals,
\href{http://arxiv.org/abs/arXiv:1702.03448}
{arXiv:1702.03448 [hep-th]}.

\bibitem{JHEP0118042}
S.M. Noorbakhsh and M.H. Vahidinia,	
Extremal vanishing horizon Kerr-AdS black holes at ultraspinning limit,
\href{http://dx.doi.org/10.1007/JHEP01(2018)042}
{\JHEP \textbf{1801} (2018) 042}.

\bibitem{PRD102-044007}
D. Wu, P. Wu, H. Yu, and S.-Q. Wu,
Are ultraspinning Kerr-Sen-AdS$_4$ black holes always superentropic?,
\href{http://dx.doi.org/10.1103/PhysRevD.102.044007}
{\PRD \textbf{102}, 044007 (2020)}.

\bibitem{PRD103-044014}
D. Wu, S.-Q. Wu, P. Wu, and H. Yu,
Aspects of the dyonic Kerr-Sen-AdS4 black hole and its ultraspinning version,
\href{http://dx.doi.org/10.1103/PhysRevD.103.044014}
{\PRD \textbf{103}, 044014 (2021)}.

\bibitem{CQG36-245017}
A.A. Balushi and R.B. Mann,
Null hypersurfaces in Kerr-(A)dS spacetimes,
\href{http://dx.doi.org/10.1088/0264-9381/15/8/012}
{\CQG \textbf{36}, 245017 (2019)}.

\bibitem{CQG38-045018}
M.T.N. Imseis, A.A Balushi, and R.B. Mann,
Null hypersurfaces in Kerr-Newman-AdS black hole and super-entropic black hole spacetimes,
\href{http://dx.doi.org/10.1088/1361-6382/abd3e0}
{\CQG \textbf{38}, 045018 (2021)}.

\bibitem{PRD103-024053}
S. Noda and Y.C. Ong,
Null hypersurface caustics, closed null curves, and super entropy,
\href{http://dx.doi.org/10.1103/PhysRevD.103.024053}
{\PRD \textbf{103}, 024053 (2021)}.

\bibitem{CQG15-2289}
F. Pretorius and W. Israel,
Quasi-spherical light cones of the Kerr geometry,
\href{http://dx.doi.org/10.1088/0264-9381/15/8/012}
{\CQG \textbf{15}, 2289 (1998)}.

\bibitem{PRD101-024057}
D. Wu, P. Wu, H. Yu, and S.-Q. Wu,
Notes on the thermodynamics of superentropic AdS black holes,
\href{http://dx.doi.org/10.1103/PhysRevD.101.024057}
{\PRD \textbf{101}, 024057 (2020)}.

\bibitem{PRD59-064005}
S.W. Hawking, C.J. Hunter, and M.M. Taylor-Robinson,
Rotation and the AdS-CFT correspondence,
\href{http://dx.doi.org/10.1103/PhysRevD.59.064005}
{\PRD \textbf{59}, 064005 (1999)}.

\end{thebibliography}
\end{document}